\documentstyle[12pt]{article}

\def\op#1{\mathop{\fam0 #1}\limits}

\newcommand{\lng}{\langle}
\newcommand{\rng}{\rangle}

\newcommand{\beq}{\begin{equation}}
\newcommand{\eeq}{\end{equation}}
\newcommand{\ben}{\begin{eqnarray}}
\newcommand{\een}{\end{eqnarray}}
\newcommand{\be}{\begin{eqnarray*}}
\newcommand{\ee}{\end{eqnarray*}}
\newcommand{\bea}{\begin{eqalph}}
\newcommand{\eea}{\end{eqalph}}

\newcommand{\cH}{{\cal H}}

\newcommand{\bR}{{\bf R}}

\newcommand{\al}{\alpha}

\newcommand{\dl}{\delta}
\newcommand{\la}{\lambda}

\newcommand{\f}{\phi}

\newcommand{\ve}{\varepsilon}

\newcommand{\dr}{\partial}

\let\ssection=\section
\renewcommand{\section}{\setcounter{equation}{0}\ssection}

\newcounter{eqalph}
\newcounter{equationa}
\newcounter{remark}
\newcounter{example}
\newcounter{theorem}
\newcounter{proposition}
\newcounter{lemma}
\newcounter{corollary}
\newcounter{definition}
\newenvironment{eqalph}{\stepcounter{equation}
\setcounter{equationa}{\value{equation}}
\setcounter{equation}{0}

\begin{eqnarray}}{\end{eqnarray}\setcounter{equation}{\value{equationa}}}

\def\theremark{\arabic{remark}}
\def\therexample{\arabic{remark}}

\def\thedefinition{\arabic{definition}}

\newenvironment{rem}{\refstepcounter{remark}\medskip\noindent{\it Remark
\theremark:}}{\medskip}

\newenvironment{prop}{\refstepcounter{definition} \medskip
\noindent{\it Proposition \thedefinition:}}{\medskip}

\begin{document}
\hbox{}

{\parindent=0pt

{\large\bf On quantum evolution as a parallel transport}
\bigskip

{\sc G. Sardanashvily}\footnote{Electronic mail:
sard@grav.phys.msu.su; sard@campus.unicam.it}

{\sl Department of Theoretical Physics, 
Moscow State University, 117234 Moscow, Russia}
\bigskip\bigskip

{\small

{\bf Abstract}. Time-dependent quantum evolution is described by an algebraic
connection on a $C^\infty(\bR)$-module of sections of a $C^*$-algebra (or
Hilbert) fibre bundle.}
}
\bigskip\bigskip

In classical time-dependent mechanics, solutions of the Hamilton equations 
can be seen as integral sections of a Hamiltonian connection, i.e., evolution
in classical Hamiltonian mechanics is described as a parallel transport along
time \cite{book98,book00,sard98,rbook98}. Following 
\cite{asor,dre,iliev,peresh,uhl}, we will treat 
evolution of a quantum time-dependent system as a parallel transport
\cite{book00,rbook00}. 

It should be emphasized that, in quantum mechanics, 
a time plays the role of a classical parameter. Indeed, 
all relations between
operators in quantum mechanics are simultaneous, while a computation of a mean
value of an operator in a quantum state  does not imply an integration over a
time. It follows that, at each moment, we have a quantum system, but these
quantum systems are different at different instants.
Although they may be isomorphic
to each other.

Recall that, in the framework of  algebraic quantum theory, a quantum
system is characterized by a
$C^*$-algebra
$A$ and a positive (hence, continuous) form $\f$ on $A$ which defines the
 representation $\pi_\f$ of
$A$ in a Hilbert space $E_\f$ with a cyclic vector $\xi_\f$ such that 
\be
\f(a)=\lng \pi_\f(a)\xi_\f|\xi_\f\rng, \qquad \forall a\in A,
\ee
(see, e.g., \cite{brat,dixm,rbook99}). One
says that
$\f(a)$ is a mean value of the operator $a$ in the state $\xi_\f$. 

Therefore, to describe quantum evolution,  one should assign a
$C^*$-algebra
$A_t$ to each point
$t\in\bR$, and treat
$A_t$ as a quantum system at the instant $t$. Thus, we have a family of
instantaneous $C^*$-algebras $A_t$, parametrised by the time axis $\bR$.
Let us suppose that all
$C^*$-algebras $A_t$ are isomorphic to each other and to some unital
$C^*$-algebra
$A$. Moreover, let they make up a locally trivial smooth Banach
fibre bundle
$P\to\bR$ with the typical fibre $A$, whose transition functions are
automorphisms of the
$C^*$-algebra $A$. 
Smooth sections  $\al$ of the $C^*$-algebra  bundle $P\to \bR$
constitute an involutive algebra with respect to the fibrewise operations.
This is also a module
$P(\bR)$ over the ring $C^\infty(\bR)$ of real smooth functions on $\bR$. 

In quantum theory, one usually follows the notion of an algebraic connection
on modules and sheaves \cite{book00,kosz60,rbook00}.
According to this notion, a connection $\nabla$  on the
$C^\infty(\bR)$-algebra
$P(\bR)$ assigns to the standard vector field $\dr_t$ on $\bR$ a
derivation of this algebra
\beq
\nabla_t\in {\rm Der}\, (P(\bR))
\label{+330}
\eeq
which obeys the Leibniz rule
\be
\nabla_t (f\al)=\dr_tf\al+ f\nabla_t \al, \qquad
\al\in
P(\bR),
\qquad f\in C^\infty(\bR).
\ee
The fibre bundle $P\to \bR$ is obviously trivial, though it has no
canonical trivialization in general.  Given its trivialization
$P=\bR\times A$, the derivation
$\nabla_t$ (\ref{+330}) reads
\beq
\nabla_t(\al) =[\dr_t - \dl(t)](\al),
\label{+331} 
\eeq
where $\dl(t)$ at each $t\in\bR$ are derivations of the $C^*$-algebra $A$,
such that
\be
\dl_t(ab)= \dl_t(a)b + a\dl_t(b), \qquad \dl_t(a^*)=\dl_t(a)^*. 
\ee

We say that a section $\al(t)$ of the fibre bundle $P\to\bR$ is an
integral section of the connection (\ref{+331}) if 
\beq
\nabla_t(\al) =[\dr_t - \dl(t)](\al)=0.
\label{+333}
\eeq
One can think of the equation (\ref{+333}) as being the  Heisenberg
equation describing quantum evolution. An
integral section $\al(t)$ of the connection $\nabla$ is a solution of this
equation. We also say that $\al(t)$ is a geodesic curve in $A$.

In particular, let the
derivations
$\dl(t)=\dl$ be the same for all
$t\in\bR$. If $\dl$ is a generator of a 1-parameter group $g_r$ of
automorphisms of the algebra $A$, then for any $a\in A$, the curve 
\beq
\al(t)=g_t(a), \qquad t\in\bR, \label{+340}
\eeq   
in $A$ is a solution of the Heisenberg equation (\ref{+333}). 

There are certain conditions in order that a derivation
$\dl$ of a unital $C^*$-algebra to define a 1-parameter group of its
automorphisms \cite{brat75,pow,rbook99}.

\begin{rem}
Let $V$ be a Banach space. An operator $a$ in $V$ is said to be bounded
if there is $\la\in \bR$ such that
\be
||av||\leq \la||v||, \qquad \forall v\in V.
\ee
The algebra $B(V)$ of bounded operators in a Banach space $V$ is a Banach
algebra with respect to the norm
\be
||a||=\op{\rm sup}_{||v||=1}||av||.
\ee
It is provided with the corresponding norm topology, called the  norm
operator topology.  Another topology in
$B(V)$, referred to in the sequel, is the strong operator topology.
It is given by the following family of
open neighbourhoods of
$0\in B(V)$:
\be
U^\ve_v=\{a\in B(V)\,:\, ||av|| <\ve\,\}, \qquad \forall v\in V,
\qquad \forall \ve>0.
\ee
 One also introduces weak, ultra-strong and ultra-weak
topologies in $B(V)$ \cite{dixm,rbook99}. Note that the
$s$-topology in
\cite{asor} is the ultra-strong topology in the terminology of \cite{dixm}.
The algebra $B(V)$ is a topological algebra only  with respect to the norm
operator topology because the morphism 
\be
B(V)\times B(V)\ni (a,b)\mapsto ab\in B(V)
\ee
 is continuous only in
this topology. The norm operator topology is finer than the other above
mentioned operator topologies.
\end{rem}

\begin{prop} \label{+348} 
If $\dl$ is a bounded derivation of a $C^*$-algebra $A$, then it is a
generator of a 1-parameter group 
\be
g_r=\exp[r\dl], \qquad r\in\bR,
\ee 
of automorphisms of $A$, and {\sl vice versa}. This group is
continuous in a norm topology in ${\rm Aut}\,(A)$. Furthermore, for any
representation
$\pi$ of
$A$ in a Hilbert space
$E_\pi$, there exists a self-adjoint bounded operator $\cH$ in $E_\pi$ such
that
\beq
\pi(\dl(a))= i[\cH,\pi(a)],\quad
\pi(g_r(a))=e^{ir\cH}\pi(a)e^{-ir\cH}, \quad \forall a\in A, \quad r\in
\bR. \label{+341}
\eeq
\end{prop}

Note that, if the domain of definition $D(\dl)$ of a derivation $\dl$
coincides with
$A$, this derivation is bounded. It follows that, by definition, the
derivations
$\dl_t$, $t\in\bR$, in the connection $\nabla_t$ (\ref{+331}) are bounded.
It follows that, if a quantum system is conservative, i.e., $\dl_t=\dl$ are the
same for all
$t\in\bR$, the Heisenberg equation (\ref{+333}) has a solution (\ref{+340})
through any point of $\bR\times A$ in accordance with Proposition
\ref{+348}.  Proposition \ref{+348} also states that the description
of evolution of a quantum conservative system in terms of the Heisenberg
equation and that based on the Shr\"odinger equation are equivalent.

However, non-trivial bounded derivations of a $C^*$-algebra do not
necessarily exist. Moreover, if a curve $g_r$ is continuous in
${\rm Aut}\,(A)$ with respect to the norm operator topology, it implies that
the curve
$g_r(a)$ for any $a\in A$ is continuous in the 
$C^*$-algebra
$A$, but the converse is not true. At the same time, a curve 
$g_r$ is continuous in ${\rm Aut}\,(A)$ with
respect to the strong operator topology in ${\rm Aut}\,(A)$ if and only if the
curve
$g_r(a)$ for any $a\in A$ is continuous in $A$. 

By this reason, we are also
interested in strong-continuous 1-parameter groups of automorphisms of
$C^*$-algebras. We refer the reader to \cite{brat75,pow,rbook99} for 
the sufficient conditions which a derivation $\dl$ should satisfy in order to
be a generator of such a group. Note only that $\dl$ has a dense domain of
definition $D(\dl)$ in $A$, and it is not bounded on $D(\dl)$. If 
$\dl$ is bounded on $D(\dl)$, then $\dl$ is extended uniquely to a bounded
derivation of $A$. It follows that, for a strong-continuous
1-parameter group of automorphisms of $A$, the connection $\nabla_t$
(\ref{+331}) is not defined on the whole algebra $P(\bR)$. In this case,
we deal with a generalized connection which is given by 
operators of a parallel transport  whose generators are not
well-defined.  It may also happen that a representation
$\pi$ of the $C^*$-algebra $A$ does not carry out the representation 
(\ref{+341}) of a strong-continuous 1-parameter group $g_r$ of automorphisms
of $A$ by unitary operators. Therefore, quantum evolution given by
strong-continuous 1-parameter groups of automorphisms need not  be
described by the Shr\"odinger equation in general.

Turn now to the time-dependent Heisenberg equation (\ref{+333}). We require
that, for all $t\in \bR$, the derivations $\dl(t)$ are generators of
strong-continuous 1-parameter groups of automorphisms of a $C^*$-algebra.
Then the operator of a parallel transport in $A$ with respect to the
connection $\nabla_t$ (\ref{+331}) over the segment $[0,t]$ can be given by 
the product integral (time-dependent exponent)
\beq
G_t=T\exp\left[ \op\int_0^t \dl(t')dt'\right]. \label{+344}
\eeq
Hence, for any $a\in A$, we have a solution 
\be
\al(t)=G_t(a),\qquad t\in \bR^+,
\ee
of the
Heisenberg equation (\ref{+333}).

Let now
all $C^*$-algebras
$A_t$ of instantaneous quantum systems be isomorphic to the von Neumann
algebra
$B(E)$ of bounded operators in some Hilbert space
$E$. Then we come to quantum evolution phrased in terms of the Shr\"odinger
equation. Let us consider a locally trivial fibre bundle $\Pi\to \bR$
with the typical fibre
$E$ and smooth transition functions. 
Smooth sections of the fibre bundle $\Pi\to \bR$ constitute a module
$\Pi(\bR)$ over the ring $C^\infty(\bR)$ of real functions on $\bR$. 
A connection $\nabla$  on
$\Pi(\bR)$ assigns to the standard vector field $\dr_t$ on
$\bR$ a first order differential operator 
\beq
\nabla_t\in
{\rm Diff}_1^\to(\Pi(\bR),\Pi(\bR))
\label{+345}
\eeq
which obeys the Leibniz rule
\be
\nabla_t (f\psi)= \dr_tf\psi+ f\nabla_t \psi, \qquad
\psi\in
\Pi(\bR),
\qquad f\in C^\infty(\bR).
\ee
Let us choose a trivialization $\Pi=\bR\times E$. Then the operator
$\nabla_t$ (\ref{+345}) reads
\beq
\nabla_t(\psi) =(\dr_t - i\cH(t)) \psi,
\label{+346} 
\eeq
where $\cH(t)$ at all $t\in\bR$ are bounded self-adjoint operators in $E$.

Note that every bounded
self-adjoint operator
$\cH$ in a Hilbert space
$E$ defines the bounded derivation 
\beq
\dl(a)=i[\cH,a], \qquad a\in B(E), \label{+346'}
\eeq
of the algebra $B(E)$. Conversely, every bounded derivation of $B(E)$ is
internal, i.e., takes the form (\ref{+346'}) where $\cH$ is a self-adjoint
element of $B(E)$. Therefore, the operators $\cH(t)$ in the expression
(\ref{+346}) are necessarily bounded and self-adjoint.

We say that a section $\psi(t)$ of the fibre bundle $\Pi\to\bR$ is an
integral section of the connection $\nabla_t$ (\ref{+346}) if it satisfies
the equation
\beq
\nabla_t(\psi) =(\dr_t - i\cH(t)) \psi=0. \label{+349}
\eeq
 One can
think of this equation as being the Shr\"odinger equation
for the Hamiltonian $\cH(t)$.

In particular, let a quantum system be conservative, i.e.,
the Hamiltonian $\cH(t)=\cH$ in the Shr\"odinger equation is independent of
time. Then, for any point $y\in E$, we obtain the
solution
\be
\psi(t)= e^{it\cH}y, \qquad t\in \bR, 
\ee
of the conservative Shr\"odinger equation. If the Shr\"odinger equation
(\ref{+349}) is not conservative, the 
operator of a parallel transport in $E$ with respect to the
connection $\nabla_t$ (\ref{+346}) over the segment $[0,t]$ can be given by 
the time-ordered exponent
\beq
G_t=T\exp\left[ i\op\int_0^t \cH(t')dt'\right]. \label{+351}
\eeq
Then, for any $y\in E$, we obtain a solution 
\beq
\psi(t)=G_ty, \qquad t\in \bR^+, \label{+355}
\eeq
of the
Shr\"odinger equation
(\ref{+349}).

Note that the operator $G_t$ (\ref{+351}) is an element of the
group
$U(E)$ of unitary operators in the Hilbert space $E$. This is a real
infinite-dimensional Lie group with respect to the norm operator
topology, whose Lie algebra is the real algebra of all anti-self-adjoint
bounded operators
$i\cH$ in $E$ with respect to the bracket $[i\cH, i\cH']$. The operator $G_t$
(\ref{+351}), by construction, obeys the equation
\beq
\dr_t G_t- i\cH G_t=0. \label{+352}
\eeq
This equation is invariant under right multiplications $G_t\mapsto G_tg$,
$\forall g\in U(E)$. Therefore, $G_t$ can be seen as the operator of a parallel
transport in the trivial principal bundle $\bR\times U(E)$. Accordingly,
the operator (\ref{+344}) can be regarded as the operator of a parallel
transport in the trivial group bundle $\bR\times {\rm Aut}\,(A)$, where the
group
${\rm Aut}\,(A)$ acts on itself by the adjoint representation.

It should be emphasized that the 1-parameter group $G_t$ defined by the
equation (\ref{+352}) is continuous with respect to  the norm operator
topology in
$U(E)$. Turn to the case when the curve
$G_t$ is continuous with respect to the strong operator topology in $B(E)$.
Then the curves 
$\psi(t)=G_ty$, $y\in E$, are also continuous, but not necessarily
differentiable in $E$. Accordingly, a Hamiltonian $\cH(t)$ in the Shr\"odinger
equation (\ref{+349}) is not bounded. Since the group $U(E)$ is not a
topological group with respect to the strong operator topology, the product
$\bR\times U(E)$ is neither principal nor smooth bundle. Therefore, the
conventional notion of a connection is not applied to this fibre bundle. At
the same time, one can introduce a generalized connection defined in terms of
parallel transport curves and operators, but not their generators \cite{asor}.

Note that the description of quantum evolution as a
parallel transport in a principal bundle can be extended to quantum systems
depending on a set of classical time-dependent parameters in order to
explain the Berry's phase phenomenon \cite{book00,rbook00,berry}.


\begin{thebibliography}{aaaa}

\bibitem{asor} M.Asorey, J.Cari\~nena and M.Paramion, Quantum evolution as a
parallel transport, {\sl J. Math. Phys.} {\bf 23} (1982) 1451.

\bibitem{brat75} O.Bratelli and D.Robinson, Unbounded derivations of
$C^*$-algebras, {\sl Commun. Math. Phys.} {\bf 42} (1975) 253; {\bf 46}
(1976) 11.

\bibitem{brat} O.Bratelli and D.Robinson, {\sl Operator Algebras and Quantum
Statistical Mechanics, Vol.1,} (Springer-Verlag, Berlin, 1979). 

\bibitem{dixm} J.Dixmier, {\sl $C^*$-Algebras} (North-Holland, Amsterdam,
1977).

\bibitem{dre} W.Drechler and Ph.Tuckey, On quantum and parallel transport in
a Hilbert bundle over space-time, {\sl Class. Quant. Grav.} {\bf 13} (1996)
611.

\bibitem{iliev} B.Iliev, Quantum mechanics from a geometric-observer's
viewpoint, {\sl J. Phys. A} {\bf 31} (1997) 1297 [quant-ph/0004041].

\bibitem{kosz60} J.Koszul, {\sl Lectures on Fibre Bundles and Differential
Geometry} (Tata University, Bombay, 1960). 


\bibitem{book98} L.Mangiarotti and G.Sardanashvily, {\sl Gauge
Mechanics} (World Scientific, Singapore, 1998).

\bibitem{book00} L.Mangiarotti and G.Sardanashvily, {\sl Connections in
Classical and Quantum Field Theory} (World Scientific, Singapore, 2000).

\bibitem{peresh} P.Pereshogin and P.Pronin, Geometrical treatment of
nonholonomic phase in quantum mechanics and applications, {\sl Int. J. Theor.
Phys.} {\bf 32} (1993) 219.

\bibitem{pow} R.Powers and S.Sakai, Unbounded derivations in operator
algebras, {\sl J. Funct. Anal.} {\bf 19} (1975) 81.

\bibitem{sard98} G.Sardanashvily, Hamiltonian time-dependent mechanics,
{\sl J. Math. Phys.} {\bf 39} (1998) 2714.

\bibitem{rbook98} G.Sardanashvily, {\sl Contemporary Methods in Field Theory:
2. Geometry and Classical Mechanics} (URSS Publ., Moscow, 1998) (in Russ).

\bibitem{rbook99} G.Sardanashvily, {\sl Contemporary Methods in Field Theory:
3. Algebraic Quantum Theory} (URSS Publ., Moscow, 1999) (in Russ).

\bibitem{rbook00} G.Sardanashvily, {\sl Contemporary Methods in Field Theory:
4. Geometry and Quantum Fields} (URSS Publ., Moscow, 2000) (in Russ).

\bibitem{berry} G.Sardanashvily, Quantum mechanics with time-dependent
parameters, E-print: quant-ph/0004005.

\bibitem{uhl} A.Uhlmann, A gauge field governing parallel transport along
mixed states, {\sl Lett. Math. Phys.} {\bf 21} (1991) 229.

\end{thebibliography}
\end{document}